\documentstyle[12pt,citesort,epsfig]{elsart} 
\newcommand{\be}[1]{\begin{equation} \label{(#1)}} 
\newcommand{\ee}{\end{equation}}  
\newcommand{\ba}[1]{\begin{eqnarray} \label{(#1)}} 
\newcommand{\ea}{\end{eqnarray}} 
\newcommand{\nn}{\nonumber}

\def\Lfv{$L_f\hspace{-0.95em}/\ \ \ $}

\def\m{$\mu^--e^-$}
\begin{document}
\begin{frontmatter}

\title{Vector mesons in nuclear $\mu^--e^-$ conversion} 
\author[Tuebingen]{Amand Faessler},  
\author[Tuebingen]{Th. Gutsche},  
\author[Valparaiso]{Sergey Kovalenko},
\author[Tuebingen]{V.E. Lyubovitskij}\footnote{On leave of absence from
Department of Physics, Tomsk State University, 634050 Tomsk, Russia},
\author[Valparaiso]{Ivan Schmidt}, 
\author[Tuebingen]{F. \v Simkovic}\footnote{On  leave of absence from
Department of Nuclear Physics, Comenius University,
Mlynsk\'a dolina F1, SK--842 15 Bratislava, Slovakia}  
\address[Tuebingen]{Institut f\"ur Theoretische Physik, Universit\"at
T\"ubingen, Auf der Morgenstelle 14, D-72076 T\"ubingen, Germany}
\address[Valparaiso]{Departamento de F\'\i sica, Universidad 
T\'ecnica Federico Santa Mar\'\i a, Casilla 110-V, 
Valpara\'\i so, Chile}

\date{\today}
 
\maketitle 
 
\vskip.5cm 

\begin{abstract}
We study nuclear $\mu^--e^-$ conversion in the general framework of an 
effective Lagrangian approach without referring to any specific 
realization of the physics beyond the standard model (SM) responsible 
for lepton flavor violation (\Lfv). 
We show that vector meson exchange between lepton and nucleon currents 
plays an important role in this process. A new issue of this mechanism 
is the presence of the strange quark vector current contribution 
induced by the $\phi$ meson.
This allows us to extract new limits on the \Lfv lepton-quark effective
couplings from the existing experimental data.

\vskip .3cm
 

\end{abstract}
\end{frontmatter}

\newpage 

\section{Introduction}

Muon-to-electron ($\mu^--e^-$) conversion in nuclei 
\begin{equation} 
\mu^- + (A,Z) \longrightarrow  e^- \,+\,(A,Z)^\ast 
\label{I.1} 
\end{equation} 
is a lepton flavor violating (\Lfv) process forbidden in the 
Standard Model (SM). It is commonly recognized as one of the most 
sensitive probes of lepton flavor violation and of related physics 
beyond the SM (for reviews, 
see~\cite{Kosmas:ch,Marciano:conf,Mue-probe}).

The present experimental situation on \m conversion is as follows. 
There is one running experiment, SINDRUM II~\cite{Honecker:zf}, 
and two planned ones, MECO~\cite{Molzon,MECO} and PRIME~\cite{PRIME}. 
The SINDRUM II experiment at PSI~\cite{Honecker:zf} with $^{48}$Ti 
as stopping target has established the best upper bound on the 
branching ratio~\cite{Honecker:zf}  
\begin{eqnarray}\label{Ti} 
&&R_{\mu e}^{Ti} = \frac{\Gamma(\mu^- + {}^{48}Ti\rightarrow e^- + 
{}^{48}Ti)} {\Gamma(\mu^- + {}^{48}Ti 
\rightarrow \nu_{\mu} + ^{48}Sc)} \leq 6.1\times 10^{-13}\ , \ \ \ 
\mbox{(90\% C.L.)} \,.  
\end{eqnarray}
The MECO experiment with $^{27}$Al is going to start soon at 
Brookhaven~\cite{MECO}. The sensitivity of this experiment is expected 
to be at the level of 
$R_{\mu e}^{Al}\leq 2\times 10^{-17}$~\cite{MECO}.   
The PSI experiment is also running with the very heavy nucleus 
$^{197}$Au aiming to improve the previous limit by the same 
experiment~\cite{Honecker:zf,Vintz} up to 
$R_{\mu e}^{Au}\leq 6 \times 10^{-13}$. 
The proposed new experiment PRIME (Tokyo)~\cite{PRIME} is going to 
utilize $^{48}$Ti as stopping target with an expected sensitivity 
of $R_{\mu e}^{Ti} \leq  10^{-18}$. 

These experimental limits would allow to set stringent limits on 
mechanisms of \m conversion and the underlying theories of \Lfv. In the 
literature various mechanisms beyond the SM have been studied 
(see~\cite{Kosmas:ch,Marciano:conf,Mue-probe} and references therein). 
They can be classified as photonic and non-photonic, that is with and 
without photon exchange between the lepton and nuclear vertices, 
respectively. These two categories of mechanisms differ significantly 
from each other in various respects. In fact, they receive  
different contributions from the new physics and also require different 
treatments of the effects of the nucleon and the nuclear structure. 
Latter aspect is, in particular, attributed to the fact that the two 
mechanisms operate at different distances and, therefore, involve 
different details of the nucleon and nuclear structure. 

In this paper we focus on the non-photonic mechanisms of \m \, 
conversion. The generic effect of physics beyond the SM in \m \,  
conversion can be described by an effective Lagrangian with all 
possible 4-fermion quark-lepton interactions consistent with Lorentz 
covariance and gauge symmetry. Here, our interest focuses on the vector 
interactions which receive a contribution from vector meson exchange. 
We will show that the vector meson contribution to the \m \, 
conversion rate is important. 

\section{General Framework}

We start with the 4-fermion effective Lagrangian describing the 
non-photonic $\mu^- - e^-$ conversion at the quark level. The most 
general form of this Lagrangian has been derived in 
Ref.~\cite{Kosmas:2001mv}. Here we present only those terms which 
contribute to the coherent $\mu^- - e^-$ conversion:
\begin{eqnarray} 
{\cal L}_{eff}^{lq}\ =\  \frac{1}{\Lambda_{LFV}^2}  
\left[(\eta_{VV}^{q} j_{\mu}^V\ + \eta_{AV}^{q} 
j_{\mu}^A )J_{q}^{V\mu} + 
(\eta_{SS}^{q} j^S\ + \eta_{PS}^{q} j^P\ )J_{q}^{S}\right], 
\label{eff-q}
\end{eqnarray}
where $\Lambda_{LFV}$ is the characteristic high energy scale of 
lepton flavor violation attributed to new physics. The summation runs 
over all the quark species $q= \{u,d,s,b,c,t\}$.  
Lepton and quark currents are defined as:  
$j_{\mu}^V = \bar e \gamma_{\mu} \mu\,, \,\,\,  
j_{\mu}^A = \bar e \gamma_{\mu} \gamma_{5} \mu\,, \,\,\,  
j^S = \bar e \ \mu$ , \mbox{$j^P = \bar e \gamma_{5} \mu $,} 
\mbox{$J_{q}^{V\mu} = \bar q \gamma^{\mu} q$}\,,$
J_{q}^{S} = \bar q \ q \,.$ 
The \Lfv parameters $\eta^{q}$ in Eq.~(\ref{eff-q}) 
depend on a concrete \Lfv model. 

The next step is the derivation of a Lagrangian in terms of effective 
nucleon fields which is equivalent to the quark level 
Lagrangian~(\ref{eff-q}). First, we write down the lepton-nucleon \Lfv 
Lagrangian of the coherent $\mu^- - e^-$ conversion in a general 
Lorentz covariant form with the isospin structure of the \m transition 
operator~\cite{Kosmas:2001mv}: 
\begin{eqnarray} 
\hspace*{-.5cm} 
{\cal L}_{eff}^{lN} =   \frac{1}{\Lambda_{LFV}^2} 
\left[j_{\mu}^a (\alpha_{aV}^{(0)} J^{V\mu \, (0)} + 
\alpha_{aV}^{(3)} J^{V\mu \, (3)}) + j^b (\alpha_{bS}^{(0)} 
J^{S \, (0)} + \alpha_{bS}^{(3)} J^{S \, (3)})\right]\,, 
\label{eff-N} 
\end{eqnarray} 
where the summation runs over the double indices $a = V,A$ and 
$b = S,P$. The isoscalar $J^{(0)}$ and isovector $J^{(3)}$ 
nucleon currents are defined as 
$J^{V\mu \, (k)} \, = \, \bar N \, \gamma^\mu \, \tau^k \,  N\,, 
\hspace*{.21cm} J^{S \, (k)}  \, = \, \bar N \, \tau^k \, N\,,$
where $N$ is the nucleon isospin doublet, $ k = 0,3 $ and 
$\tau_0\equiv\hat I$. 
This Lagrangian is supposed to be generated by the one of 
Eq.~(\ref{eff-q}) and, therefore, must correspond to the same  
order $1/\Lambda_{LFV}^{2}$ in inverse powers of the \Lfv scale. 
The Lagrangian~(\ref{eff-N}) is the basis 
for the derivation of the nuclear transition operators. 

Now one needs to relate the lepton-nucleon \Lfv parameters $\alpha$  
in Eq.~(\ref{eff-N}) to the more fundamental lepton-quark \Lfv 
parameters $\eta$ in Eq.~(\ref{eff-q}). This implies a certain 
hadronization prescription which specifies the way in which the effect 
of quarks is simulated by hadrons. In the absence of a true theory of 
hadronization we rely on some reasonable assumptions and models. 

There are basically two possibilities for the hadronization mechanism. 
The first one is a direct embedding of the quark currents into the 
nucleon (Fig.1a), which we call direct nucleon mechanism (DNM).  
The second possibility is a two stages process (Fig.1b). First, the   
quark currents are embedded into the interpolating meson fields which
then interact with the nucleon currents. We call this possibility 
meson-exchange mechanism (MEM). 
In general one expects all the mechanisms to contribute 
to the coupling constants $\alpha$ in Eq.~(\ref{eff-N}). However, 
at present the relative amplitudes of each mechanism are unknown. 
In view of this problem one may try to understand the importance of 
a specific mechanism, assuming for simplicity, that only this 
mechanism is operative and estimating its contribution to the process 
in question. We follow this procedure for the case of $\mu^- - e^-$ 
conversion and consider separately the contributions of the direct 
nucleon mechanism $\alpha_{[N]}$ and the meson-exchange one 
$\alpha_{[MN]}$ to the couplings of the Lagrangian (\ref{eff-N}).  

In the present paper we concentrate on the meson-exchange mechanism.
The contribution of the direct 
nucleon mechanism has been derived in Ref.~\cite{Kosmas:2001mv}. 
Here we present only the results of Ref.~\cite{Kosmas:2001mv} relevant 
for our analysis which are the couplings of the vector nucleon currents 
in Eq.~(\ref{eff-N}): 
\begin{eqnarray} 
\label{alpha}
\hspace*{-.65cm}
\alpha_{aV[N]}^{(3)} = \frac{1}{2}(\eta_{aV}^{u} - \eta_{aV}^{d}) 
(G_{V}^{u} - G_{V}^{d})\,, 
\,\,\,  
\alpha_{aV[N]}^{(0)}  = \frac{1}{2}(\eta_{aV}^{u} + \eta_{aV}^{d}) 
(G_{V}^{u} + G_{V}^{d})\,,
\end{eqnarray} 
where $a=V,A$. The nucleon form factors $G_{V}^{q}$ define 
the strong isospin symmetric normalization of 
the quark current matrix elements between 
nucleon states: 
\begin{eqnarray}\label{mat-el1} 
\langle p|\bar{u}\ \gamma_{\mu}\ u|p\rangle &=&  
G_{V}^{u} \bar{p}\ \gamma_{\mu}\ p,  \ \ \  
\langle n|\bar{d}\ \gamma_{\mu}\ d|n\rangle = 
G_{V}^{u} \bar{n}\ \gamma_{\mu}\ n,\\ \nn 
\langle p|\bar{d}\ \gamma_{\mu}\ d|p\rangle &=&  
G_{V}^{d} \bar{p}\ \gamma_{\mu}\ p, \ \ \  
\langle n|\bar{u}\ \gamma_{\mu}\ u|n\rangle = 
G_{V}^{d} \bar{n}\ \gamma_{\mu}\ n. \\ \nn 
\end{eqnarray}
Since the maximal momentum transfer $q$ in $\mu^- -e^-$ conversion is 
much smaller than the typical scale of the nucleon structure 
we can safely neglect the $q^2$-dependence of the nucleon form factors 
$G_{V}^{q}$. At $q^2=0$ these form factors are equal to the total 
number of the corresponding species of quarks in the nucleon and, 
therefore, 
$G_{V}^{u}=2,\   G_{V}^{d}=1$, while the form factors corresponding to 
$s, c, b, t$ quarks are equal to zero. This is the reason why they 
do not contribute to the couplings of the vector nucleon current in 
Eq.~(\ref{alpha}). In the next section we will show that the vector 
meson exchange may drastically modify this situation and introduce 
the contribution of strange quarks into these couplings.

\section{Vector Meson Contribution} 

Now let us turn to the contributions of the meson-exchange mechanism to 
the couplings $\alpha$ of the lepton-nucleon Lagrangian (\ref{eff-N}).  
The mesons which can contribute to this mechanism are the unflavored 
vector and scalar ones. Since the case of the scalar meson candidate 
$f_0(600)$ is still quite uncertain~\cite{Hagiwara:fs} we do not study 
its contribution. Thus, we are left with the vector mesons. 
The lightest of them, giving the dominant contributions, are the 
isovector $\rho(770)$ and the two isoscalar $\omega(782), 
\ \phi(1020)$ mesons. 
We use ideal singlet-octet mixing for the quark content of the
$\omega$ and $\phi$ mesons~\cite{Hagiwara:fs}: 
$\omega = (u \bar u + d \bar d)/\sqrt{2}$ and 
$\phi = - s \bar s\,.$ 

First, we derive the \Lfv lepton-meson effective Lagrangian in terms of 
the interpolating $\rho^0, \ \omega$ and $\phi$ fields retaining all 
the interactions consistent with Lorentz and electromagnetic gauge 
invariance. It can be written as: 
\begin{eqnarray}\label{eff-LV}
{\cal L}_{eff}^{lV}\ &=& \  \frac{\Lambda_H^2}{\Lambda_{LFV}^2}  
\biggl[ \, \biggl\{ (\xi_V^{\rho} j_{\mu}^V\ + \xi_A^{\rho} 
j_{\mu}^A\ )\rho^{0\ \mu} + (\xi_V^{\omega}  j_{\mu}^V\ 
+ \xi_A^{\omega} j_{\mu}^A\ )\omega^{\mu} + \\ \nn 
 &+& (\xi_V^{\phi}  j_{\mu}^V\ + \xi_A^{\phi} 
j_{\mu}^A\ )\phi^{\mu} \biggr\}  + \frac{1}{\Lambda_H^2}
\biggl\{ \xi_V^{\rho(2)} j_{\mu}^V\ 
\partial^{\mu}\partial^{\nu} \rho^0_{\nu} + ... \biggr\} + ... \biggr], 
\end{eqnarray}  
with the unknown dimensionless coefficients $\xi$ to be determined 
from the had\-ro\-ni\-za\-ti\-on prescription. Since this Lagrangian is 
supposed to be generated by the quark-lepton Lagrangian (\ref{eff-q})  
all its terms have the same suppression $\Lambda_{LFV}^{-2}$ with 
respect to the large \Lfv scale $\Lambda_{LFV}$. Another scale in the 
problem is the hadronic scale $\Lambda_H \sim 1$ GeV which adjusts the 
physical dimensions of the terms in  Eq.~(\ref{eff-LV}). Typical 
momenta involved in $\mu^- - e^-$ conversion are $q \sim m_{\mu}$, 
where $m_{\mu}$ is the muon mass. Thus, from naive dimensional counting 
one expects that the contribution of the derivative terms to 
$\mu^- - e^-$ conversion is suppressed by a factor 
$(m_{\mu}/\Lambda_H)^2\sim 10^{-2}$ in comparison to the contribution 
of the non-derivative terms. Therefore, at this step in 
Eq.~(\ref{eff-LV}) we retain only the dominant non-derivative terms. 
However, it is worth noting that a true hadronization theory, 
yet non-existing, may forbid such terms so that the expansion in 
Eq.~(\ref{eff-LV}) starts from the derivative terms. In a forthcoming 
paper~\cite{Long_paper} we are going to demonstrate how this happens 
in a particular model of hadronization and what is the phenomenological 
impact of this situation.  

In order to relate the parameters $\xi$ of the 
Lagrangian (\ref{eff-LV}) with the ``fundamental" parameters $\eta$ of 
the quark-lepton Lagrangian (\ref{eff-q}) we use an approximate method 
based on the standard on-mass-shell matching 
condition~\cite{Faessler:1996ph}  
\begin{equation}\label{match} 
\langle \mu^+ \, e^-|{\cal L}_{eff}^{lq}|V\rangle \approx 
\langle \mu^+ \, e^-|{\cal L}_{eff}^{lV}|V \rangle ,  
\end{equation} 
with $|V= \rho, \omega, \phi \rangle$ corresponding to vector meson 
states on mass-shell. We solve equation (\ref{match}) using the 
well-known quark current matrix elements 
\begin{eqnarray}\label{mat-el2} 
&&\langle 0|\bar u \, \gamma_\mu \, u|\rho^0(p,\epsilon)\rangle \, = \, 
\, - \, \langle 0|\bar d \, \gamma_\mu \, d|\rho^0(p,\epsilon)\rangle 
\, = \, m_{\rho}^2 \, f_{\rho} \, \epsilon_\mu(p), \nonumber\\ 
&&\langle 0|\bar{u}\ \gamma_{\mu}\ u|\omega(p,\epsilon)\rangle = 
\langle 0|\bar{d}\ \gamma_{\mu}\ d|\omega(p,\epsilon)\rangle 
\, = \, 3 \, m_{\omega}^2 \, f_{\omega} \, \epsilon_\mu(p)\,,\\ 
&&\langle 0|\bar{s}\ \gamma_{\mu}\ s|\phi(p,\epsilon)\rangle 
\, = \, - \, 3 \, m_\phi^2 \, f_{\phi} \, \epsilon_\mu(p) \,.  
\nonumber
\end{eqnarray} 
Here $p$, $m_V$, $f_{V}$ and $\epsilon_{\mu}$ are the 
vector-meson four-momentum, mass, $V\to\gamma$ decay constant and 
the polarization state vector, respectively. The quark operators 
in Eq.~(\ref{mat-el2}) are taken at $x=0$. The coupling constants 
$f_{V}$ are determined from the $V\to e^+ e^-$ decay width: 
$\Gamma(V \to e^+ e^-) \, = \, (4\pi/3) \, \alpha^2 \, 
f_{V}^2 \, m_V$\,, where $\alpha$ is the fine-structure constant.
The current central values of the meson couplings $f_{V}$ 
and masses $m_V$ are~\cite{Hagiwara:fs}: 
\begin{eqnarray}\label{constants}  
&&\hspace*{.15cm}f_{\rho}   = 0.2 \,, \hspace*{1.7cm} 
  f_{\omega} = 0.059  \,, \hspace*{1.5cm} 
  f_{\phi}   = 0.074  \,\,, \\   
&&m_{\rho}   = 771.1 \,\,  \mbox{MeV}, \,\,\,
  m_{\omega} = 782.57 \,\, \mbox{MeV}, \,\,\,  
  m_{\phi}   = 1019.456 \,\, \mbox{MeV}\,. \nonumber
\end{eqnarray} 
Solving Eq.~(\ref{match}) with the help of Eqs.~(\ref{mat-el2}), 
we obtain the desired expressions for the coefficients $\xi$ of the 
lepton-meson Lagrangian (\ref{eff-LV}) in terms of generic \Lfv  
parameters $\eta$ of the lepton-quark effective Lagrangian 
Eq.~(\ref{eff-q}): 
\begin{eqnarray}
&&\xi_a^{\rho} \, = \, \left(\frac{m_{\rho}}{\Lambda_H}\right)^2 
f_{\rho} \,(\eta_{a V}^{u} - \eta_{a V}^{d})\,, \,\, 
\xi_a^{\omega} \, = \, 3 \left(\frac{m_{\omega}}{\Lambda_H}\right)^2  
\, f_{\omega} \, (\eta_{a V}^{u} + \eta_{a V}^{d}) \,, \,\,\\ \nn
&&\xi_a^{\phi} \, = \, - 3 \left(\frac{m_{\phi}}{\Lambda_H}\right)^2  
\, f_{\phi} \, \eta_{a V}^{s} \, ,
\end{eqnarray} 
where $a = V, A$. 

Now we derive the vector meson exchange contributions 
to the couplings of the effective Lagrangian in 
Eq.~(\ref{eff-N}) expressed on the nucleon level.
To this end we introduce the effective Lagrangian 
describing the interaction of nucleons with vector 
mesons~\cite{Weinberg:de,Mergell:1995bf,Kubis:2000zd}: 
\begin{eqnarray}\label{MN} 
{\cal L}_{VN} \, = \, \frac{1}{2} \, 
\bar{N}\gamma^{\mu}\left[ g_{_{\rho NN}} \, \vec{\rho}_{\mu} \, 
\vec{\tau} \,  + \, g_{_{\omega NN}} \, \omega_{\mu} \, + \, 
g_{_{\phi NN}} \, 
\phi_{\mu}\right] N\,. 
\end{eqnarray} 
In this Lagrangian we neglected the derivative terms, irrelevant for 
coherent $\mu^- - e^-$ conversion. For the meson-nucleon couplings 
$g_{VNN}$ we use numerical values taken from an updated dispersive 
analysis~\cite{Mergell:1995bf} 
\begin{eqnarray}\label{VN-couplings}
g_{_{\rho NN}}= 4.0\,, \,\, 
g_{_{\omega NN}} = 41.8\,, \,\, 
g_{_{\phi NN}}= - 18.3\,. 
\end{eqnarray} 
Substituting the values of the $g_{VNN}$ constants from 
Eq.~(\ref{VN-couplings}) into the $SU(3)$ relation~\cite{deSwart:gc} 
$g_{_{\phi NN}} \, = \, g_{_{\rho NN}} \, (\sqrt{3}/\cos\theta_V) \, 
- \, g_{_{\omega NN}} \, \tan\theta_V\,,$ we estimate the 
$\omega-\phi$ mixing angle to be $\theta_V = 32.4^{\rm o}$,  
which is close to the ideal one $\theta_V^I = 35.3^{\rm o}$, 
consistent with our initial assumption on the quark content of the 
$\omega$ and $\phi$ mesons.  

At this point the following comment is in order. The relatively large 
value of the ${\phi NN}$  coupling $g_{_{\phi NN}} = - 18.3$ 
in Eq.~(\ref{VN-couplings}) has been derived in 
Ref.~\cite{Mergell:1995bf} on the basis of the assumption on 
the "maximal" violation of the Okuba-Zweig-Iizuka (OZI) 
rule. It was also stressed in Ref.~\cite{Mergell:1995bf} that this 
value corresponds to the upper limit for the ${\phi NN}$ coupling 
which parameterizes the full spectral function in the mass region of 
$\sim$1 GeV within $\phi$ pole dominance approximation. The inclusion  
of other contributions such as the $\pi\rho$ continuum leads to 
a significant reduction of the $g_{_{\phi NN}}$ 
coupling~\cite{Meissner:1997qt}. The detailed analysis of various 
meson and baryon cloud contributions to the vector ${\phi NN}$ 
coupling results in the value 
$g_{_{\phi NN}} = - 0.24$~\cite{Meissner:1997qt}. 
In what follows, we quote this value as "physical" one. 
For completeness in our numerical analysis we consider both values 
of $g_{_{\phi NN}}$ coupling: 
\begin{eqnarray}\label{max-phys}
g_{_{\phi NN}}^{\rm phys} = - 0.24 \ \mbox{("physical")}, \ \ \    
g_{_{\phi NN}}^{\rm max}= - 18.3 \ 
\mbox{("maximal")}\,. 
\end{eqnarray} 
The vector meson-exchange contribution to the nucleon-lepton effective 
La\-gran\-gi\-an (\ref{eff-N}) arises in second order in the Lagrangian 
${\cal L}_{eff}^{lV} + {\cal L}_{VN}$  
and corresponds to the diagram in Fig. 1(b). 
We estimate this contribution 
only for the coherent $\mu^- - e^-$ conversion process. 
In this case we disregard all the derivative terms of nucleon and 
lepton fields. Neglecting the kinetic energy of the final nucleus, 
the muon binding energy and the electron mass, 
the momentum transfer squared $q^2$ to the nucleus has a constant value
$q^2 \approx - m_{\mu}^2$. 
In this approximation the vector meson propagators convert to 
$\delta$-functions leading to lepton-nucleon contact operators. 
Comparing them with the corresponding terms in the 
Lagrangian (\ref{eff-N}), we obtain for the vector meson-exchange 
contribution to the coupling constants:
\begin{eqnarray} \label{alpha-V-ex}
\alpha_{aV[MN]}^{(3)} &=& - \beta_{\rho}(\eta_{aV}^{u} - \eta_{aV}^{d}),
\ \ \  
\alpha_{aV[MN]}^{(0)} = - \beta_{\omega}(\eta_{aV}^{u} + \eta_{aV}^{d}) 
- \beta_{\phi}\eta_{aV}^{s}\,,
\end{eqnarray}
with $a=V,A$ and the coefficients 
\begin{eqnarray} \label{beta}
\beta{_{\rho}} = \frac{1}{2} 
\frac{g_{_{\rho NN}} \, f_{\rho} \, m_{\rho}^2}{m_{\rho}^2 
+ m_{\mu}^2},\ \beta{_{\omega}} = \frac{3}{2} 
\frac{g_{_{\omega NN}} \, f_{\omega} \, m_{\omega}^2 }
{m_{\omega}^2 + m_{\mu}^2},\ \beta{_\phi} = -\frac{3}{2} 
\frac{g_{_{\phi NN}} \, f_{\phi} \, 
m_{\phi}^2 }{m_{\phi}^2 + m_{\mu}^2}. 
\end{eqnarray}
Substituting the values of the meson coupling constants and masses 
from Eqs.~(\ref{constants}) and (\ref{VN-couplings}), and including 
the two different options for the $g_{_{\phi NN}}$ constant 
(\ref{max-phys}), we obtain for these coefficients 
\begin{eqnarray} \label{beta-num} 
\beta{_{\rho}} = 0.39\,, \,\,\, 
\beta{_{\omega}} = 3.63\,, \,\,\, 
\beta{_{\phi}}^{\rm phys} = 0.03\,, \,\,\, 
\beta{_{\phi}}^{\rm max} = 2.0 \,\,. 
\end{eqnarray} 
A new issue of the vector meson contribution (\ref{alpha-V-ex})
is the presence of the strange quark vector current contribution 
associated with the LFV parameter $\eta_{aV}^{s}$, absent in the 
direct nucleon mechanism as it follows from Eq.~(\ref{alpha}). 
This opens up the possibility of deriving new limits on this parameter 
from the experimental data on $\mu^- - e^-$ conversion. Another 
surprising result is that the contribution~(\ref{alpha-V-ex}) of 
the meson-exchange mechanism is comparable to the 
contribution~(\ref{alpha}) of the direct nucleon mechanism.

\section{Constraints on \Lfv parameters from $\mu^- - e^-$ conversion}

>From the Lagrangian~(\ref{eff-N}), following the standard procedure, 
one can derive the formula for the branching ratio of coherent 
$\mu^- - e^-$ conversion. To leading order in the non-relativistic 
reduction the branching ratio takes the form~\cite{Kosmas:ch}
\begin{equation} 
R_{\mu e^-}^{coh} \ = \  
\frac{{\cal Q}} {2 \pi \Lambda_{LFV}^4} \  \   
\frac{p_e E_e \ ({\cal M}_p + {\cal M}_n)^2 } 
{ \Gamma ({\mu^-\to capture}) } 
\, , 
\label{Rme}
\end{equation} 
where $p_e, E_e$ are 3-momentum and energy of the outgoing electron,  
${{\cal M}}_{p,n}$ are the nuclear $\mu^- - e^-$ 
transition matrix elements. 
The factor ${\cal Q}$ has the form~\cite{Kosmas:2001mv} 
\begin{eqnarray}
\hspace*{-1cm}
{\cal Q} &=& |\alpha_{VV}^{(0)}+\alpha_{VV}^{(3)}\ \phi|^2 +
|\alpha_{AV}^{(0)}+\alpha_{AV}^{(3)} \phi|^2 + 
|\alpha_{SS}^{(0)}+\alpha_{SS}^{(3)} \phi|^2 + 
|\alpha_{PS}^{(0)} + \alpha_{PS}^{(3)} \phi|^2 
\nonumber  \\ 
\hspace*{-1cm} 
&+& 2{\rm Re}\{(\alpha_{VV}^{(0)}+\alpha_{VV}^{(3)} 
\phi)(\alpha_{SS}^{(0)}+ \alpha_{SS}^{(3)} \phi)^\ast
+  (\alpha_{AV}^{(0)}+\alpha_{AV}^{(3)}\ \phi)(\alpha_{PS}^{(0)} + 
\alpha_{PS}^{(3)}\ \phi)^\ast\}\,.
\label{Rme.1} 
\end{eqnarray} 
in terms of the parameters of the lepton-nucleon effective 
Lagrangian (\ref{eff-N}) and the nuclear structure factor 
$\phi = ({\cal M}_p - {\cal M}_n)/({\cal M}_p + {\cal M}_n) \, $. 

In Refs.~\cite{Kosmas:ch,Kosmas:2001mv} it was argued that for all 
the experimentally interesting nuclei the parameter $\phi$ is small 
$\sim 10^{-2}$. Therefore the nuclear structure dependence of 
${\cal Q}$ can always be neglected except for a very special narrow 
domain in the \Lfv parameter space where 
$\alpha^{(0)}\leq \alpha^{(3)}\phi$. Another important issue of the 
smallness of the ratio $\phi$ in Eq.~(\ref{Rme.1}) is the dominance of 
the isoscalar contribution associated with the coefficients 
$\alpha^{(0)}$. For this reason the role of the $\rho$-meson exchange 
in $\mu^- - e^-$ conversion is expected to be unimportant except for 
the above mentioned very special case.    

The nuclear matrix elements ${\cal M}_{p,n}$ in Eq.~(\ref{Rme}) have 
been calculated in Refs.~\cite{Kosmas:2001mv,Faessler:pn} for the 
nuclear targets $^{27}$Al, $^{48}$Ti and $^{197}$Au. With these matrix 
elements upper limits on the \Lfv parameters have been deduced
from the experimental constraints~\cite{Kosmas:2001mv}.
In particular, for the case of the present experimental constraint 
(\ref{Ti}) it was found that 
\begin{equation}\label{alpha-lim}
\alpha_{aV}^{(0)} \left(\frac{1 \mbox{GeV}}{\Lambda_{LFV}}\right)^2 
\leq 1.2 \times 10^{-12},
\end{equation}
for the combination of the dimensionless couplings $\alpha_{aV}^{(0)}$
$(a = A,V)$ and the characteristic \Lfv scale $\Lambda_{LFV}$ in the 
effective lepton-nucleon Lagrangian (\ref{eff-N}). From the above limit 
one can deduce the individual limits on different terms determining 
the coefficients $\alpha_{aV}^{(0)}$ in the direct nucleon and 
meson-exchange mechanisms, assuming that significant cancellations 
(unnatural fine-tuning) between different terms are absent.  
In this way we derive constraints for the $\eta$ parameters of the
quark-lepton Lagrangian~(\ref{eff-q}) for the meson-exchange
mechanism (MEM). We present these limits in Table~1. For the parameter
$|\eta_{aV}^{s}|$ we derive the limits for the two different cases of
$g_{_{\phi NN}}$ coupling, given in Eq.~(\ref{max-phys}). In Table~1
we also show for comparison the limits corresponding to the direct
nucleon mechanism (DNM) derived in Ref.~\cite{Kosmas:2001mv}.
The limits presented in Table~1 show the importance of the vector
meson exchange contribution to $\mu^- - e^-$ conversion. 

\section{Summary} 

We studied nuclear $\mu^--e^-$ conversion in a general framework 
based on an effective Lagrangian without referring to any specific 
realization of the physics beyond the standard model responsible 
for lepton flavor violation. We demonstrated that the vector 
meson-exchange contribution to this process is significant.  
A new issue of the meson-exchange mechanism in comparison 
to the previously studied direct nucleon mechanism is the 
presence of the strange quark vector current contribution induced by 
the $\phi$ meson. This allowed us to extract new limits 
on the \Lfv lepton-quark effective couplings from the existing 
experimental data. 

\vspace*{.2cm}

{\bf Acknowledgments}

\noindent 
This work was supported by the DAAD under contract 
415-ALECHILE/ALE-02/21672, by the FONDECYT projects 1030244, 1030355,  
by the DFG under contracts FA67/25-3, 436 SLK 113/8 and GRK683, 
by the State of Baden-W\"{u}rt\-tem\-berg, LFSP "Low Energy Neutrinos", 
by the President grant of Russia 1743 "Scientific Schools", 
by the VEGA Grant agency of the Slovac Republic under contract 
No.~1/0249/03.

\newpage 

\begin{table}

\centerline{LIST OF TABLES}

\vspace*{.5cm} 

{\bf Table 1.} Upper bounds on the \Lfv parameters inferred from the 
SINDRUM II data on $^{48}$Ti [Eq.~(\ref{Ti})] 
corresponding to the direct nucleon mechanism (DNM)
and the meson exchange mechanism (MEM). The subscript notation is $a = V,A$. 
The value in square brackets refers to the $g_{_{\phi NN}}^{\rm max}$ value 
of $\phi NN$ coupling
presented in Eq. (\ref{max-phys}).  

\vspace*{.4cm} 

\begin{center} 
\begin{tabular}{|c|c|c|} 
\hline \hline 
Parameter & DNM &MEM \\ 
\hline 
&&    \\ 
$|\eta_{aV}^{u,d}|(1 \, \mbox{GeV}/\Lambda_{LFV})^2$ 
& $8.0\times 10^{-13}$&$3.3\times 10^{-13}$\\  
&&    \\ 
$|\eta_{aV}^{s}|(1 \, \mbox{GeV}/\Lambda_{LFV})^2$& no limits
& $4.0\times 10^{-11}$ \,\, $[6.0\times 10^{-13}]$\\ 
&&    \\ 
\hline         
\hline
\end{tabular}
\end{center} 
\end{table}

\begin{figure}[t]

\vspace*{1cm}
\centerline{LIST OF FIGURES}
\vspace*{.3cm} 

\noindent {\bf Fig.1:} 
Diagrams contributing to the nuclear $\mu^--e^-$ conversion: 
direct nucleon mechanism (1a) and 
meson-exchange mechanism (1b). 

\vspace*{.5cm}
\centering{\
\epsfig{figure=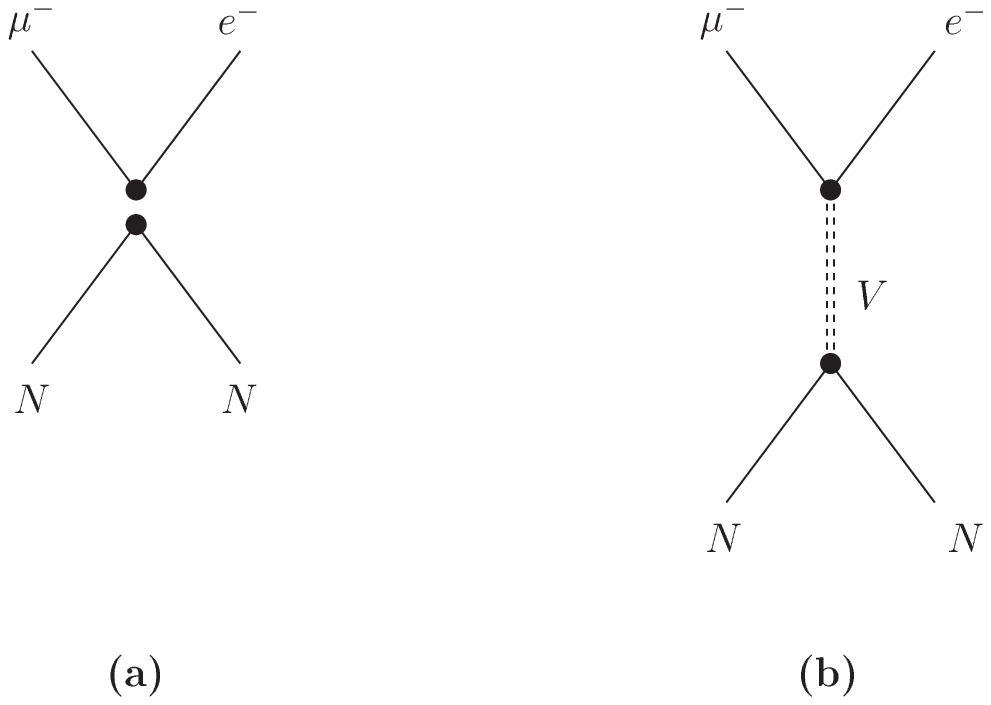,height=7cm}}

\centerline{\bf Fig.1} 
\end{figure}
\end{document}